\documentclass[doublecol]{epl2}

\title{Can scale-freeness offset delayed signal detection in neuronal \\ networks?}
\shorttitle{Can scale-freeness offset delayed signal detection in neuronal networks?}

\author{Rukiye Uzun,\inst{1} Mahmut Ozer\inst{1} \and Matja{\v z} Perc\inst{2}}
\shortauthor{Rukiye Uzun, et al.}

\institute{\inst{1}Department of Electrical and Electronics Engineering, B{\"u}lent Ecevit University, 67100 Zonguldak, Turkey\\
\inst{2}Faculty of Natural Sciences and Mathematics, University of Maribor, Koro{\v s}ka  cesta 160, SI-2000 Maribor, Slovenia}


\abstract{First spike latency following stimulus onset is of significant physiological relevance. Neurons transmit information about their inputs by transforming them into spike trains, and the timing of these spike trains is in turn crucial for effectively encoding that information. Random processes and uncertainty that underly neuronal dynamics have been shown to prolong the time towards the first response in a phenomenon dubbed noise-delayed decay. Here we study whether Hodgkin-Huxley neurons with a tunable intensity of intrinsic noise might have shorter response times to external stimuli just above threshold if placed on a scale-free network. We show that the heterogeneity of the interaction network may indeed eradicate slow responsiveness, but only if the coupling between individual neurons is sufficiently strong. Increasing the average degree also favors a fast response, but it is less effective than increasing the coupling strength. We also show that noise-delayed decay can be offset further by adjusting the frequency of the external signal, as well as by blocking a fraction of voltage-gated sodium or potassium ion channels. For certain conditions, we observe a double peak in the response time depending on the intensity of intrinsic noise, indicating competition between local and global effects on the neuronal dynamics.}

\begin{document}

\maketitle

\section{Introduction}
The dynamics of complex systems, the backbone of which are often complex interaction networks, has been the subject of intense study during recent years \cite{boccaletti_pr06, sporns_tcs04}. Seminal works on network science \cite{watts_dj_n98, barabasi_s99} have revealed that many real-world networks exhibit small-world or scale-free topological properties, and neuronal networks are by no means an exception \cite{bullmore_nrn09}. For example, Egu{\'i}luz et al. \cite{eguiluz_prt05} examined the organization of the functionally connected human brain during a resting state on a voxel scale and observed a scale-free architecture of functionally connected brain regions. Moreover, Fraiman et al. \cite{fraiman_pre09} compared networks derived from the fMRI signals of the human brain with similar networks extracted form the Ising model, and they found that near the critical temperature the two networks are similar. In fact, there exist many more evidence in support of criticality and emergent collective behavior in the dynamics of neuronal networks \cite{chialvo_np10}, the majority of which are a direct consequence of the complex interaction patterns among neurons.

Compared to large-scale neuronal networks, neurons alone are relatively simple excitable units that respond by means of stereotyped pulses, called action potentials or spikes, to extrinsic stimuli that can be provided by external excitation, by noise, or by neighboring neurons in a spatially extended system \cite{balenzuela_pre05, ma_cpl08, tang_epl12} (for comprehensive reviews see \cite{lindner_pr04, neiman_s07, sagues_rmp07}). The basis of information processing in the brain is transforming the incoming signals into neuronal excitations. These transformations are crucial for efficient encoding of information \cite{lochmann_newjp08}, and the processing capacity of neurons is directly related to the nature of spike trains as a code \cite{schneidman_anyps00}. Spike trains can encode information via timing (temporal coding) \cite{abeles_91} or via mean firing rates (rate coding) \cite{adrian_jp26, steveninck_prslsb98}. Since there exist evidence that rate coding would in many situations be inefficient and unreliable compared to temporal coding \cite{tuckwell_b05}, the focus is shifting towards the later. In the context of temporal coding, the timing of the first spike is of particular relevance, as it typically carries a greater amount of information about the incoming stimulus than subsequent spikes \cite{panzeri_n01, thorpe_90}. It is within this context and with this motivation that Pankratova et al. \cite{pankratova_epjb05, pankratova_pla05} analyzed the impact of external noise on the timing of signal detection in FitzHugh-Nagumo and Hodgkin-Huxley neurons. The reported results revealed that the first spike latency is inversely proportional to the noise strength, and that it can be minimized by a proper driving frequency of the external stimulus. However, if the noise intensity exceeds a threshold, the latency again begins decreasing. The phenomenon that thus an intermediate noise intensity yields the slowest responsiveness of an individual neuron was described as noise-delayed decay (NDD), and subsequent studies followed up by examining the relevance of internal noise through neuronal membrane patches with stochastic channels \cite{ozer_jtb09}, synaptic background activity \cite{uzuntarla_epjb12}, as well as temperature variations \cite{uzun_ksej13}. Ozer and Graham \cite{ozer_epjb08} examined the NDD in dependence on the network activity by varying the membrane time constant of a single cell. They showed that NDD emerges for small values of the time constant, thus indicating high network activity, and that it vanishes for large values. In addition to these studies at the level of a single cell, in \cite{ozer_pla08} the authors analyzed the first spike latency on small-world neuronal networks, and showed that it exists for small coupling strengths. The NDD phenomenon has thus already received substantial attention in terms of the relevance of internal noise through neuronal membrane patches with stochastic channels and synaptic background activity.

Neurons are inherently noisy, and in general it is believed that noise has a destructive impact on the effectiveness of neuronal information processing, although it can also aid the detection of weak signals through stochastic resonance \cite{gammaitoni_rmp98}. Voltage-gated ion channels embedded in neuronal membrane are one of the major sources of noise due to their random transitions between conducting and nonconducting states \cite{steinmetz_jcn00}. The intensity of channel noise is related to the number of active ion channels that participate in the generation of spikes \cite{schneidman_nc98}, and assessing the impact of the number of active ion channels is therefore important, especially to uncover the role of specific ion channel noise on neuronal firing. In this context, neurotoxins such as tetraethlyhammonium and tetrodotoxin are used in experiments to reduce the number of working ion channels \cite{hille_01}. In particular, by means of a fine-tuned adminstration of these toxins a certain fraction of potassium or sodium ion channels can be disabled or blocked. The relevance of ion channel noise can also be studied by means of computational models. It is known, for example, that the regularity of spontaneous spike trains can be reduced or enhanced by blocking or poisoning some fraction of sodium or potassium ion channels \cite{schmid_pa04, schmid_pb04}, and also that channel blocking can enhance the collective spiking regularity of bi-directionally coupled \cite{gong_sxsbc08} and small-world neuronal networks \cite{ozer_epl09}. Recently, the development, propagation or robustness of spiral waves observed in the cortex of brain has been addressed via ion channel poisoning \cite{wu_cnsns13, ma_ctp13, huang_jbs13}.

In this letter, we build on these previous advances to determine the role of interaction networks by delayed signal detection, in particular by the timing of first spikes in scale-free coupled Hodgkin-Huxley neurons. Thereby, we employ a model for the stochastic behavior of the voltage-gated ion channels embedded in the membrane patch, where the channel noise intensity depends on the membrane area. We focus on the relevance of the strength of coupling, the average degree of individual neurons, as well as on the frequency of the external signal and the fraction of blocked voltage-gated sodium or potassium ion channels. We thus deliver a comprehensive study that reveals under which conditions the heterogeneity of neuronal networks actually enhances responsiveness, which factors that may reduce spike latency, and ultimately under which conditions the temporal coding might be optimal. The main results and conclusions are presented in subsequent sections, while first we describe the mathematical model and other details of the setup in greater detail.

\section{Mathematical model and setup}
In the network, the dynamics of each neuron is described by the Hodgkin and Huxley \cite{hh_jp52} model, according to which the time evolution of the membrane potential for coupled neurons in the presence of an external signal $f(t)=A \sin(\omega t)$ is given as follows:
{\setlength\arraycolsep{0.1em}\begin{eqnarray}
C_{m}{\rm d}V_{i}/{\rm d}t=&&G_{Na}(m_i,h_i)(V_{Na}-V_i)+G_K(n_i)(V_K-V_i)\nonumber\\*
&&+G_L(V_L-V_i)+\sum_{j}\varepsilon_{ij}(V_j-V_i)+f(t)
\end{eqnarray}}
In Eq.~(1) $\varepsilon^{i,j} = \varepsilon$ is the coupling strength if neuron $i$ is coupled to neuron $j$, while otherwise $\varepsilon^{i,j} = 0$, $C_m=1{\rm\mu Fcm^{2}}$ is the membrane capacity, and $G_{Na}$, $G_K$ and $G_L$ represent sodium, potassium and leakage conductances, respectively. Moreover, $V_{Na}=115{\rm mV}$, $V_{K}=-12{\rm mV}$ and $V_{L}=10.6{\rm mV}$ are the reversal potentials for the sodium, potassium and leakage channels. The leakage conductance is set constant at $G_{L}=0.3{\rm mScm^{-2}}$, while the sodium and potassium conductances change dynamically according to \cite{schmid_pa04, schmid_pb04, ozer_epl09, ozer_jtb09}:
\begin{equation}
G_{Na}(m_i,h_i)=g^{max}_{Na}x_{Na}m^{3}_{i}h_{i},\
G_{K}(n_i)=g^{max}_{K}x_{K}n^{4}_{i}
\end{equation}
In Eq.~(2) $g^{max}_{Na}=120{\rm mScm^{-2}}$ and $g^{max}_{K}=36{\rm mScm^{-2}}$ are the maximal sodium and potassium conductances, respectively. Moreover, $m$ and $h$ denote the activation and inactivation gating variables for the sodium channel, respectively, whereas the potassium channel includes an activation gating variable $n$. We also introduce two scaling factors, $x_{Na}$ and $x_K$, which are the fractions of non-blocked ion channels comparing to the total number of sodium ($N_{Na}$) or potassium ($N_K$) ion channels within the patch area, respectively \cite{schmid_pa04, schmid_pb04, ozer_epl09, ozer_jtb09}. These scaling factors are confined to the unit interval.

\begin{figure}
\centerline{\scalebox{0.465}[0.465]{\includegraphics{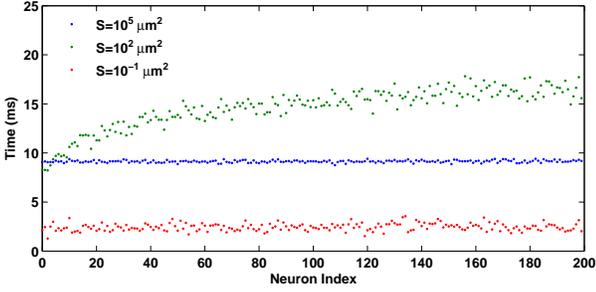}}}
\caption{Spatiotemporal distribution of first spikes produced by $200$ scale-free coupled Hodgkin-Huxley neurons driven by a supra-threshold periodic stimulus $f(t)$ with frequency $\nu = 20 {\rm Hz}$ and amplitude $A=4{\rm\mu A / cm^{2}}$ for different membrane areas $S$, as indicated in the figure legend. An intermediate intensity of internal noise at $S= 100 {\rm \mu m^{2}}$ induces maximal delays, which however are less pronounced by high-degree neurons (low neuron index $i$) than by low degree neurons. This hints towards the fact that the heterogeneity of the interaction network might play a key role in mitigating prolonged first spike latency. The employed coupling strength is $\varepsilon=0.01$.}
\label{basics}
\end{figure}

Activation and inactivation gating variables, $m_i$, $n_i$ and $h_i$, change over time in response to the membrane potential following first-order differential equations, but only in the limit of very large cell sizes. However, since the population of ion channels is finite, the stochastic behavior of voltage-gated ion channels must be taken into consideration. To account for this, we use the algorithm proposed by Fox \cite{fox_bj97}. Thus, variables of stochastic gating dynamics are described with the corresponding Langevin generalization \cite{fox_bj97}:
\begin{equation}
{\rm d}x_{i}/{\rm d}t=\alpha_x(1-x_i)-\beta_x x_i+\xi_{x_{i}}(t),\
x_i=m_i,n_i,h_i
\end{equation}
where $\alpha_x$ and $\beta_x$ are rate functions for the gating variable $x_i$. The probabilistic nature of the channels appears as a source of noise $\xi_{x_{i}}(t)$ in Eq.~(3), which is an independent zero mean Gaussian noise whose autocorrelation function is given as follows \cite{schmid_pa04, schmid_pb04, ozer_epl09, ozer_jtb09}:
\begin{equation}
\langle \xi_m(t) \xi_m(t') \rangle = \frac{2 \alpha_m \beta_m}{N_{Na}x_{Na}(\alpha_m + \beta_m)} \delta(t-t')
\end{equation}
\begin{equation}
\langle \xi_h(t) \xi_h(t') \rangle = \frac{2 \alpha_h \beta_h}{N_{Na}x_{Na}(\alpha_h + \beta_h)} \delta(t-t')
\end{equation}
\begin{equation}
\langle \xi_n(t) \xi_n(t') \rangle = \frac{2 \alpha_n \beta_n}{N_{K}x_{K}(\alpha_n + \beta_n)} \delta(t-t')
\end{equation}
where the factors, $x_{Na}$ and $x_K$, are used again to disregard the blocked channels, which do not contribute to the intrinsic channel noise. Given the assumption of homogeneous sodium and potassium ion channel densities, channel numbers are calculated via $N_{Na}=\rho_{Na}S$, $N_{K}=\rho_{K}S$ where $\rho_{Na}=60{\rm\mu m^{-2}}$ and $\rho_{K}=18{\rm\mu m^{-2}}$ are the sodium and potassium channel densities, respectively, whereas $S$ represents the total membrane area, done previously in \cite{schmid_pa04, schmid_pb04, ozer_jtb09, ozer_epl09}. Equations (4-6) define that the intrinsic noise level is inversely proportional to the number of ion channels in the membrane area.

As the interaction network describing the connections between the neurons we use the scale-free network generated via growth and preferential attachment as proposed by Barab{\'a}si and Albert \cite{barabasi_s99}. Here growth implies that the numbers of connected neurons increases with time, while preferential attachment means that new neurons are more likely to connect with existing neurons that already have a large number of connections to other neurons. Typically, we use networks with an average degree $k_{avg}$ consisting of $N=200$ neurons, although we have verified that the presented results are independent of the system size.

We quantify the response of the network by means of the mean latency $\Lambda$, which measures the average time neurons in the network need to produce the first spike in response to the external signal. Accordingly, $\Lambda=N^{-1}\sum_i t_i$, where $t_i$ is the first response time of neuron $i$ that is recorded as soon as the membrane potential $V_i$ crosses the $20 {\rm mV}$ threshold upwardly for the first time. We also determine the second moment of $t_i$ as $\Psi=\sqrt{N^{-1}\sum_i t_i^2-\Lambda^2}$, which represents the so-called temporal jitter. Final values of $\Lambda$ and $\Psi$ presented below are averages over up to 100 independent runs conducted for each set of parameter values to ensure appropriate statistical accuracy with respect to the scale-free network generation and stochastic simulations.

\section{Results}

To begin with, we note that for the external signal $f(t)=A \sin(\omega t)$ we use $A=4{\rm\mu A / cm^{2}}$ and $\nu = 20 {\rm Hz}$, where $\omega=2 \pi \nu$, which according to \cite{pankratova_epjb05} is just above the firing threshold of $\nu = 16 {\rm Hz}$ at this particular amplitude. Intrinsic noise does thus not play the role of the main excitatory agent, but rather it masks the deterministic external signal $f(t)$. The task of the neuronal network is to detect and respond to $f(t)$ as soon as possible.

\begin{figure}
\centerline{\scalebox{0.48}[0.48]{\includegraphics{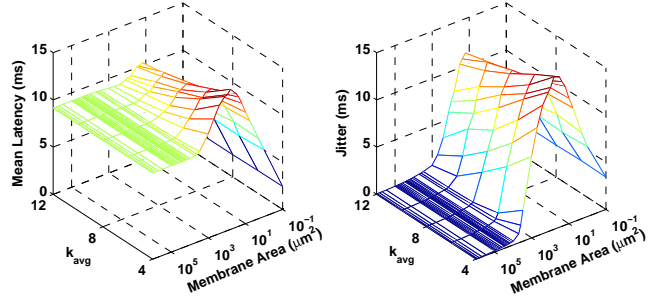}}}
\caption{Mean latency $\Lambda$ (left) and jitter $\Psi$ (right) in dependence on the average degree $k_{avg}$ and the membrane area $S$, as obtained for $\nu = 20 {\rm Hz}$, $A=4{\rm\mu A /cm^{2}}$ and $\varepsilon=0.01$. It can be observed that there exists an intermediate internal noise intensity at which both $\Lambda$ and $\Psi$ are maximal. Although the peak values decrease with increasing $k_{avg}$ and also shift towards slightly larger values of $S$, the relevance of the average degree of individual neurons is fairly marginal.}
\label{degree}
\end{figure}

In Fig.~\ref{basics}, we first show the spatiotemporal distribution of neuronal firings for three characteristics values of $S$. It can be observed that the first spike latency is maximal at an intermediate total membrane area equalling $S= 100 {\rm \mu m^{2}}$
(green dots), while for $S= 0.1 {\rm \mu m^{2}}$ (red dots) and $S= 10^5 {\rm \mu m^{2}}$ (blue dots) the first response times are, at least overall, significantly shorter. This is the hallmark property of noise-delayed decay. An intermediate intensity of intrinsic noise, here directly regulated by the membrane area $S$, maximally delays the response of the neuronal network, and in so doing compromises its ability to detect stimulus onset as well as to effectively encode information transmitted via $f(t)$. Importantly, however, from Fig.~\ref{basics} one can also observe that, especially for the most damaging value of $S= 100 {\rm \mu m^{2}}$, the response times differ significantly depending on the neuron number $i$. According to the employed growth and preferential attachment algorithm \cite{barabasi_s99}, low-index neurons are the oldest and thus also the most interconnected neurons within the network. These neurons are able to detect and respond to $f(t)$ much faster than high-index neurons. The green line in Fig.~\ref{basics} has a persistent upward trend towards larger $t_i$ as $i$ increases from $0$ towards $N-1$, and the trend is particularly strong for the first $\approx 20$ neurons. According to the scale-free degree distribution, these few high-degree neurons hold contact with the majority of other neurons in the network, and it is likely that the auxiliary input coming from all these other neurons helps the high-degree neurons to detect the deterministic external signal faster than low-degree neurons. This may in turn constitute a mechanism by means of which noise-delayed decay in scale-free neuronal networks could be avoided altogether, and in what follows, we elaborate on this perspective in more detail.

\begin{figure}
\centerline{\scalebox{0.48}[0.48]{\includegraphics{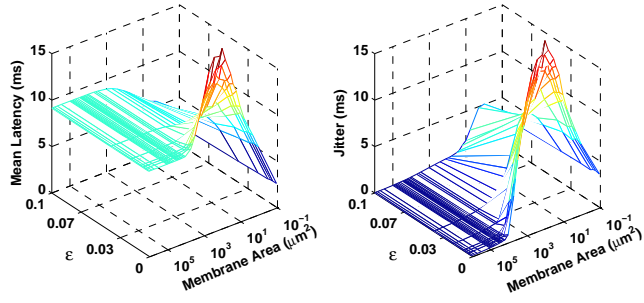}}}
\caption{Mean latency $\Lambda$ (left) and jitter $\Psi$ (right) in dependence on the coupling strength $\varepsilon$ and the membrane area $S$, as obtained for $\nu = 20 {\rm Hz}$, $A=4{\rm\mu A /cm^{2}}$ and $k_{avg}=4$. As in Fig.~\ref{degree}, it can be observed that there exists an intermediate internal noise intensity at which both $\Lambda$ and $\Psi$ are maximal. As $\varepsilon$ increases, however, the peak values decrease significantly, and at sufficiently strong coupling the signature of noise-delayed decay appears to altogether vanish. This indicates that the coupling strength plays a more pivotal role than average degree in mitigating delayed first-spike onset in neuronal networks.}
\label{coupling}
\end{figure}

Figure~\ref{degree} shows how the average degree $k_{avg}$ and the intrinsic noise strength (regulated via $S$) affect the mean latency $\Lambda$ and the temporal jitter $\Psi$. It can be observed that increasing $k_{avg}$ does slightly better the responsiveness of the network, but also that the positive effect is quite marginal. Moreover, the intermediate value of $S$ evoking the most delayed response is hardly affected and remains bounded between $S \approx 50 {\rm \mu m^{2}}$ (for large $k_{avg}$) and $S \approx 100 {\rm \mu m^{2}}$ (for low $k_{avg}$). Thus, we conclude that the average degree of the scale-free network does not play a key role in lessening noise-delayed decay.

\begin{figure}
\centerline{\scalebox{0.48}[0.48]{\includegraphics{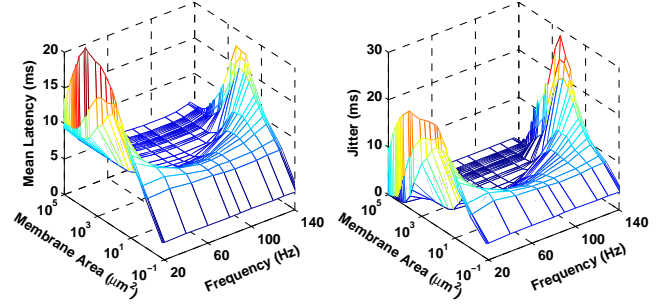}}}
\caption{Mean latency $\Lambda$ (left) and jitter $\Psi$ (right) in dependence on the frequency of the external signal and the membrane area $S$, as obtained for $\varepsilon=0.01$, $k_{avg}=4$ and $A=4{\rm\mu A /cm^{2}}$. It can be observed that low and high frequencies are significantly more elusive to detection than intermediate frequencies of the external signal. The latter yield a clear decrease in both $\Lambda$ and $\Psi$, although an intermediate values of $S$ still proves to be the most damaging to early detection. These results are largely robust to variations of $\varepsilon$ and $k_{avg}$, and they also agree with preceding observations made on individual neurons.}
\label{freq}
\end{figure}

Results presented in Fig.~\ref{coupling} are more promising, where indeed a significant drop in both the maximal mean latency $\Lambda$ and the maximal temporal jitter $\Psi$ can be observed as the coupling strength $\varepsilon$ increases. More precisely, $\Lambda$ drops by a factor of two and $\Psi$ by a factor of three as $\varepsilon$ goes from $0.001$ to $0.1$. In addition, when approaching $\varepsilon=0.1$ the mean latency $\Lambda$ no longer displays a resonant-like dependence on the total membrane area $S$, and for the temporal jitter $\Psi$ the bell-shaped form fades significantly as well. Although the most damaging values of $S$ remain, as when increasing $k_{avg}$ (see Fig.~\ref{degree}), relatively unaffected, the results in Fig.~\ref{coupling} nevertheless do lend credible support to the notion that the network structure could be crucial for mitigating noise-delayed decay.

\begin{figure}
\centerline{\scalebox{0.48}[0.48]{\includegraphics{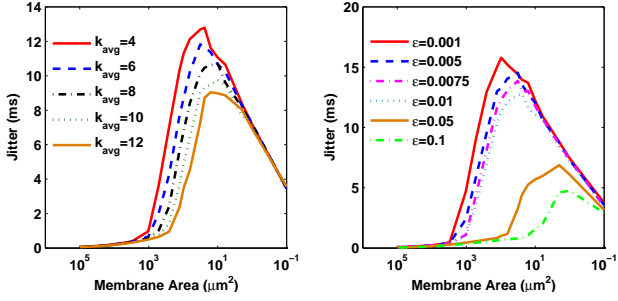}}}
\caption{Temporal jitter in dependence on the membrane area $S$, as obtained for different average degrees $k_{avg}$ (left) and coupling strengths $\varepsilon$ (right) in the presence of potassium channel blocking ($x_K=0.95$ and $x_{Na}=1$). Similarly as concluded from results presented in Figs.~\ref{degree} and \ref{coupling}, in case of potassium channel blocking too the coupling strength plays a much more significant role than the average degree in ensuring high responsiveness of the neuronal network.}
\label{potassium}
\end{figure}

In order to determine the conditions under which this may apply more accurately, we first change the properties of the external signal $s(t)$, in particular its frequency $\nu$. Figure~\ref{freq} shows how $\Lambda$ and $\Psi$ vary in dependence on $\nu$ and $S$. Interestingly, the network structure does not affect the frequency range in which delayed responses were recorded before at the individual neuron level. Referring to Fig.~1 in \cite{pankratova_epjb05}, we find that for the amplitude $A=4{\rm\mu A / cm^{2}}$ only a rather narrow interval centering on $\nu \approx 90 {\rm Hz}$ does not evoke noise-delayed decay. This is in full agreement with results presented in Fig.~\ref{freq}, thus indicating that with regards to the properties of the external signal, the mitigation of noise-delayed decay by means of network structure can take full reference from the response of an individual neuron. As we will show in what follows, this paves the way for an intricate interplay between local (individual-neurons based) and global (network-based) effects that affect first spike latency, ultimately giving rise to a double resonance-like dependence of $\Lambda$ and $\Psi$ on $S$ in case of selective sodium or potassium channel blocking.

\begin{figure}
\centerline{\scalebox {0.48}[0.48]{\includegraphics{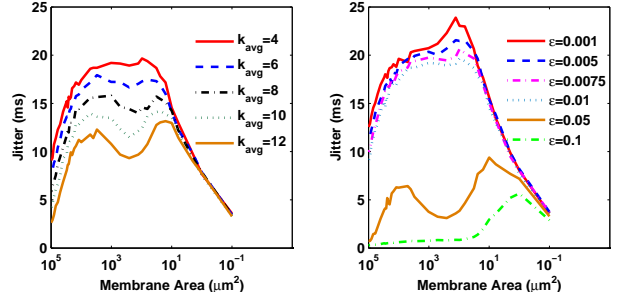}}}
\caption{Temporal jitter in dependence on the membrane area $S$, as obtained for different average degrees $k_{avg}$ (left) and coupling strengths $\varepsilon$ (right) in the presence of sodium channel blocking ($x_{Na}=0.95$ and $x_K=1$). Here too the coupling strength plays a key role in reducing the signature of noise-delayed decay, while the impact of average degree is much more subtle. The increase of both $\varepsilon$ and $k_{avg}$, however, introduces a peculiar double-peak structure in dependence on $S$, which indicates competition between local and global effects affecting the delayed responsiveness of the neuronal network.}
\label{sodium}
\end{figure}

The impact of potassium channel blocking is presented in Fig.~\ref{potassium}, where the earlier results in Figs~\ref{degree} and \ref{coupling} are reviewed for $x_K=0.95$ while keeping $x_{Na}=1$. Here we focus solely on the temporal jitter, as the second moment is expectedly more sensitive to the change in $t_i$, although we note that the examination of the mean latency $\Lambda$ would lead to identical conclusions. From Fig.~\ref{potassium} it follows that results in the absence of channel blocking apply also for the case of potassium channel blocking. In particular, while the increase in the average degree $k_{avg}$ plays a side role at best, the increase in the coupling strength $\varepsilon$ has the potency to significantly dampen noise-delayed decay. In addition, if potassium channels are blocked, the increase in $\varepsilon$ is also accompanied by a strong shift in the most damaging value of $S$. While for small coupling strengths $S \approx 100 {\rm \mu m^{2}}$ is most effective in delaying neuronal response to $f(t)$, at high coupling strengths this shifts by two orders of magnitude to $S \approx 1 {\rm \mu m^{2}}$. The fact that a much smaller total membrane area is needed to evoke the maximally delayed response indicates that blocking potassium channels may significantly increase the robustness of neuronal networks to noisy disturbances.

Sodium channel blocking promises further insights, in that for certain $k_{avg}$ and $\varepsilon$ values there emerges a double peak in $\Psi$ in dependence on $S$, as evidenced in Fig.~\ref{sodium}. While the peak around $S \approx 100 {\rm \mu m^{2}}$ was highlighted before for $x_K=x_{Na}=1$ as well as for $x_K=0.95$ and $x_{Na}=1$, for $x_K=1$ and $x_{Na}=0.95$ there emerges a persistent second peak at $S \approx 5 {\rm \mu m^{2}}$, which is particularly pronounced at high $k_{avg}$ and high coupling strength $\varepsilon$. The very low intrinsic noise intensity that evokes the doubly noise-delayed decay, together with the need for relatively strongly coupled neurons constituting the scale-free network, suggests that an extreme sensitization to weak external stimuli sets in, which in turn very effectively masks the main signal and thus induces failure of the network to respond in a timely manner. Sodium channel blocking has before been linked to increased sensitivity of neuronal dynamics \cite{ozer_epl09}, only that here this is additionally amplified by the scale-free interaction structure. Indeed, as Fig.~\ref{basics} indicates, the high-degree neurons (low $i$ values) are by default more responsive than low-degree neurons ($i$ close $N$), which in the presence of sodium channel blocking may result in tuning-in too much even to the faintest of noisy disturbances, and thus facilitating the numbness to the actual signal that ought to be detected.

\section{Summary}
We have studied noise-delayed decay on scale-free neuronal networks subject to intrinsic noise, different frequencies of the external signal to be detected, as well as subject to separate potassium and sodium channel blocking. We have shown that the scale-free interaction structure amongst neurons has the potential to significantly shorten the response time of the entire network, which is due to the higher responsiveness of the high-degree neurons. For the mechanism to work, however, the coupling strength has to be sufficiently strong, so that the faster response of the high-degree neurons can be detected back also by the low-degree neurons in a timely manner. Ozer and Uzuntarla \cite{ozer_pla08} have arrived at qualitatively similar findings by using a different network topology, small-world neuronal network, indicating that strong coupling between neurons reduces the NDD effect regardless of the network topology. Increasing the average degree also has a positive impact, but its magnitude is only a fraction of that of the high coupling strengths. In terms of the relevance of the frequency of the external signal, we have shown that full reference can be taken from the preceding study of the responsiveness of an individual neuron. We have also demonstrated that potassium channel blocking increases the robustness of neuronal networks to noisy disturbances, while sodium channel blocking induces a doubly noise-delayed decay. We have concluded that the temporal jitter peak at high membrane area values is a consequence of the scale-free network structure that further amplifies the increased sensitivity due to sodium channel blocking, while the peak at low membrane area values is due to the inherent neuronal dynamics. Only the latter can be tamed effectively by a heterogeneous network structure as long as the coupling between the neurons is sufficiently strong and the frequency of the external signal avoids the prohibitive values set by individual neuronal dynamics, while the former can be considered as the ''price to pay'' for the aforementioned benefits. Given that first spike latency following stimulus onset -- the effective manifestation of noise-delayed decay -- is physiologically relevant in that it may prevent effective signal detection and responsiveness, and ultimately lead also to inefficient information encoding based on spike train timing, we believe the study addresses a relevant setup with potential practical ramifications. We hope this will be motivation enough for further research efforts aimed at disentangling the importance of network structure in mitigating delayed first responses of neuronal networks due to noise. In particular, although many works have focused only on electrically coupled neurons, as we have also done in our study, the role of chemical coupling \cite{connors_arn04, balenzuela_pre05, yilmaz_pa13, liu_csf14} as well as other sources of heterogeneity besides the network structure might merit particular attention in either impairing or promoting NDD.

\acknowledgments
This work was supported by the Slovenian Research Agency (Grant J1-4055).


\end{document}